\documentclass[%
 aip,
 amsmath,amssymb,
 reprint,%
]{revtex4-2}
\usepackage{graphicx}
\usepackage{dcolumn}
\usepackage{bm}

\usepackage[utf8]{inputenc}
\usepackage[T1]{fontenc}
\usepackage{mathptmx}
\usepackage{etoolbox}

\makeatletter
\def\@email#1#2{%
 \endgroup
 \patchcmd{\titleblock@produce}
  {\frontmatter@RRAPformat}
  {\frontmatter@RRAPformat{\produce@RRAP{*#1\href{mailto:#2}{#2}}}\frontmatter@RRAPformat}
  {}{}
}%
\makeatother
\begin{document}

                                                                                                                                                                                                                                                                                                                                                                                                                                                                                                                                                                                                                                                                                                                                                                                                                                                                                                                                                                                                                                                                                                                                                                                                                                                                                                                                                                                                                                                                                                                                                                                                                                                                                                                                                                                                                                                                                                                                  \title{Effect of doping on hot-carrier thermal breakdown in perforated
graphene metasurfaces}
                                                                                                                                                                                                                                                                                                                                                                                                                                                                                                                                                                                                                                                                                                                                                                                                                                                                                                                                                                                                                                                                                                                                                                                                                                                                                                                                                                                                                                                                                                                                                                                                                                                                      \author{M.~Ryzhii$^{1*}$, V.~Ryzhii$^{2,3*}$, C. Tang$^{2,3}$,
                                                                                                                                                                                                                                                                                                                                                                                                                                                                                                                                                                                                                                                                                                                                                                                                                                                                                                                                                                                                                                                                                                                                                                                                                                                                                                                                                                                                                                                                                                                                                                                                                                                                                                                                                                                                                                                                                                                           T. Otsuji$^{4,5,6}$,  and M. S. Shur$^{7,8}$
}
\address{$^1$School of Computer Science and Engineering, University of Aizu, Aizu-Wakamatsu
 965-8580,\\ Japan \\
$^2$Research Institute of Electrical Communication,~Tohoku University,~Sendai~980-8577,\\ Japan \\
$^3$Frontier Research Institute for Interdisciplinary Sciences,
Tohoku University, Sendai 980-8578,\\ Japan \\
$^4$ International Research Institute of Disaster Science, Tohoku University,
Sendai 980-8578,\\ Japan \\
$^5$Center of Excellence ENSEMBLE3, Warsaw 01-919,\\
 Poland \\
$^6$UNIPRESS Institute of High Pressure Physics of the Polish Academy of Sciences, Warsaw 02-822,\\ Poland \\
$^7$Department of Electrical,Computer, and Systems Engineering,  
Rensselaer Polytechnic Institute,~Troy,~New York~12180, USA\\
$^8$Electronics of the Future, Inc., Vienna, VA 22181-6117,\\ USA\\
*{Authors to whom correspondence should be addressed: m-ryzhii@u-aizu.ac.jp, vryzhii@gmail.com}}

\begin{abstract}
We examine the robustness of the S-shaped current–voltage characteristics associated with hot-carrier–induced electrical breakdown in perforated graphene metasurfaces (PGMs) as a function of doping. The perforation of the graphene layer forms interdigital arrays of graphene microribbons (GMRs) interconnected by graphene nanoribbon (GNR) bridges. These GNR constrictions act as energy barriers for electrons and holes emitted from the GMRs and govern the inter-GMR thermionic current under an applied bias voltage. The doping and the voltage bias establish distinct electron and hole populations in adjacent GMRs. Peltier heating of these carriers within the GMRs increases their effective temperatures, thereby enhancing the inter-GMR current. The resulting positive feedback between carrier heating and current amplification can trigger an electrothermal breakdown, transforming a superlinear current–voltage dependence into an S-shaped characteristic exhibiting negative differential resistance. The degree of electron–hole asymmetry significantly influences this positive feedback and strongly modifies the overall current-voltage response. These results provide a framework for optimizing PGM-based devices employing GMR/GNR architectures, including voltage-controlled fast switches, incandescent emitters, and terahertz bolometric detectors.
\end{abstract}

\maketitle
\section{Introduction}
The unique properties of graphene layers (GLs),~\cite{1,2,3,4} in particular, patterned into graphene micro- and nanoribbons (GMRs and GNRs) and  combined with two-dimensional van der Waals structures~\cite{5,6,7,8,9,10} 
   enable many device applications: terahertz and infrared detectors and sources, transistors  and other devices.~\cite{11,12,13,14,15,16,17,18,19,20,21,22,23,24,25,26,27}
Recently, we demonstrated that perforated graphene layers, constituting
 arrays of coplanar interdigital graphene micro- and nanoribbons (GMRs and GNRs), can exhibit specific and pronounced 
 plasmonic oscillations~\cite{28} and the  hot-carrier thermal breakdown,~\cite{29,30}  which can be used in effective effective perforated graphene metasurfaces (PGMs) based terahertz photomixers and detectors~\cite{31,32,33}
 and fast voltage-controlled switching devices.  

The physical-mathematical model used in the previous analysis of similar PGMs assumed that the pristine GL and, hence, the GMRs fabricated via the perforation procedure, are undoped.
Unintentional or intentional chemical  GL doping can lead to a substantial  non-equivalence 
of the oppositely polarized neighboring GMRs.  
As might be expected, this can result in different carrier densities and carrier emission from neighboring GMRs, as well as their effective temperatures.
In this paper, we extend our previous analysis and device model, 
providing a more comprehensive description of carrier transport and electrothermal feedback in doped PGM structures.

\begin{figure*}[t]
\centering
\includegraphics[width=12cm]{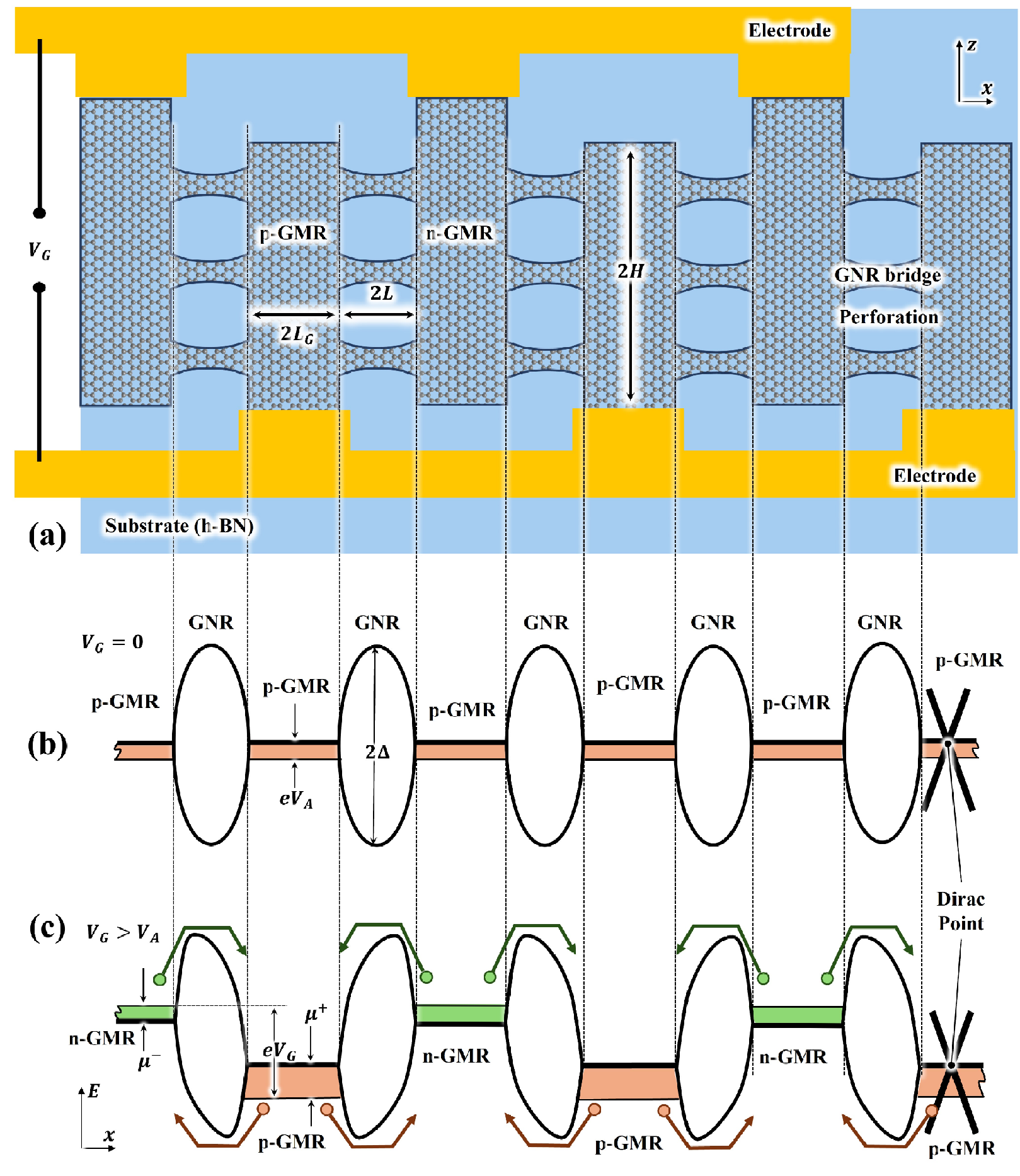}
\caption{ (a) Top  view of a PGM with  metal  contact electrodes, (b) its band diagram corresponding to cross-sections of the inter-GNR bridges at $V_G =0 $ and (c) at $V_G > V_A$.}
\label{Fig1}
\end{figure*}

\section{Main Equations of the PGM  Model}

Figure~1(a) presents the PGM based on 
  the GMR/GNR structure created by the GL perforation.
In the absence of the bias voltage, all the GMRs are filled with holes with the density close to  acceptor density $\Sigma_A$ (all the GMRs are of p-type). 
At sufficiently high
bias voltages $V_{G} > V_A$, where $V_A \propto \Sigma_A$ is the doping (acceptor) voltage,
 the 2D electron and 2D hole systems are formed
 in the neighboring GMRs so that the n-GMRs and p-GMRs alternate.
In the areas near the  perforations, the n-GMRs and p-GMRs are separated by relatively high energy barriers for the electrons and holes. 
The heights of these barriers are determined by the band alignment
between graphene and the substrate material, for example h-BN (although other substrate materials that provide a high-quality interface, such as SiC, can be used). 
In contrast, the GNRs offer relatively low energy barriers between the n-GMRs and p-GMRs. These barriers are associated with the lateral confinement  of  electron and hole motion in the GNRs
(perpendicular to the GNRs) and the pertinent quantization of their
energy spectra. Hence,   the height of such barriers is determined by the GNR minimal thickness $w$. Figures~1(b) and 1(c) show the  band diagrams without bias voltage and under a sufficiently high bias voltage $V_{G} > V_A$, respectively. 
The GMR ends are connected with the contact metal (or graphene) pads.
The GMR/GNR structures under consideration are of  $2H$ and $2L_G$ in the GMR length and width, respectively, with   $(2N-1)$  of the GNRs bridging each GMR pair. The shape of the GL perforations is assumed to be relatively smooth 
(like an oval) so that the GNR barrier can be considered as approximately parabolic.

Using Landauer-Buttiker approach,~\cite{34} the thermionic  current, $j_{GMR}$, through each GNR from one GMR to neighboring GMR can be presented as

\begin{eqnarray}\label{eq1}
j_{GNR} = j_{GNR}^{-}+  j_{GNR}^{+}
\end{eqnarray}
with 

\begin{eqnarray}\label{eq2}
j_{GNR}^{\mp} \simeq 
\frac{4e}{\pi\,\hbar}\biggl[T^{\mp}\exp\biggl(\frac{\mu^{\mp}-\Delta +\eta\,eV_G}{T^{\mp}}\biggr)\nonumber\\
- T^{\pm}\exp\biggl(\frac{-\mu^{\mp}-\Delta -\eta\,eV_G}{T^{\pm}}\biggr)\biggr]
\end{eqnarray}
being the electron (marked by "-") and hole (marked by "+") components.
Here, $T^{\pm}$ are the electron and hole effective temperatures, 
 $\mu^{\mp} \simeq \hbar\,v_W\sqrt{\pi \Sigma_G^{\mp}}$ are  the  Fermi energies in the pertinent GMRs  with $\Sigma_G^{\mp} = C_GV_G/2L_Ge \mp \Sigma_A$
(for acceptor doping with the surface density $\Sigma_A$). 
The  Fermi energies of the carriers induced in the GMRs by both acceptors and the bias voltage at not too low their  densities can be presented as $\mu^{\mp} \simeq e\sqrt{{\overline V}_G(V_G \mp V_A)}$, $V_A \simeq 2eL_G\Sigma_A/C_G$,
   ${\overline V}_G = (\pi\,C_G\hbar^2v_W^2/2e^3L_G)$ is the characteristic voltage, 
  $c_G= [(\kappa_S+1)/4\pi^2]{\overline C}_G$ is the inter-GMR capacitance,
which is determined by the geometrical parameters ratio $L_G/L$  and   the substrate dielectric constant $\kappa_S$~\cite{35}  
 (see as well Refs.~36 and 37), $v_W$ is the carrier velocity in GLs, and $\hbar$ is the Planck constant.
We assume that   the GNR barrier is lowered  by the voltage between the GMR 
so that  the barrier is equal to $\Delta -\eta\,eV_G$, where $\eta$ is determined by the shape of the GNR energy barrier.  For near trapezoidal barriers $\eta \ll 1$, whereas for   near-parabolic barriers with the top at the GNR center under consideration    $\eta \simeq 1/2$.
The latter is assumed  in the following.  

The second terms on the right sides of Eqs.~(\ref{eq2}), which are relatively small at sufficiently large $V_G$, correspond to the reverse thermionic currents - the electron current from the p-GMRs and the hole current from the n-GMRs.
Equations~(\ref{eq2}) account for possible distinctions in the Fermi energies $\mu^-$ and $\mu^+$
and in the effective temperatures $T^{-}$ and $T^{+}$ in the n- and p-GMRs.

The electrons coming from the n-GMR into the p-GMR 
increase the energy of the holes and, therefore, heat the hole system in the latter,
while the holes from the p-GMR heat the electron system in the n-GMR (the carrier heating due to the Peltier effect). 
Accounting for that the Joule heating associated with the currents along the GMRs can be weak if the conductance along them is high,
we, disregarding the latter,  use the following equations for the energy
balance in the n- and p-GMRs:

\begin{eqnarray}\label{eq3}
2HC_G(V_G \mp V_A)R^{\mp}
= (2N-1)ej_{GNR}^{\pm}V_G.
\end{eqnarray}
 This, in particular,  implies a relatively small contribution of the quantum capacitance to the net inter-GMR capacitance and a smallness of the built-in voltage. According to the latter, we preserve the quantities $\mu^{\mp}$ 
 in the exponents in Eqs.~(\ref{eq2}), disregarding them in other terms.
The validity of Eqs.~(\ref{eq3}) is limited by the bias voltage range  
${\overline V}_G,\,\mu^{\mp} \ll , V_G,  (V_G \mp V_A)$, in which one can expect
 the thermal breakdown effect explored below.

Considering the PGM structures at room temperature,
we assume that 
the carrier energy relaxation is associated  with optical phonons being the crucial mechanism (see, for example,  Refs.~38 - 41).
When the electron and hole systems are heated somewhat above room temperature,
the rate of optical phonon relaxation can saturate. This is the main reason for the hot-carrier breakdown at a certain bias voltage $V_G$ (at the threshold voltage $V_{G}^{th}$).
In this case, other energy  relaxation mechanisms can appear, determining   
the PGM structure characteristics beyond the threshold. 
For the sake of definiteness, we assume that such mechanisms are associated with the disordered-assisted electron-phonon scattering and the plasmon-assisted scattering.~\cite{42,43,44}
On the analogy with the previous studies (see, for example,~Refs.~45 and 46),
we set

\begin{eqnarray}\label{eq4}
 R^{\mp} \simeq \frac{\hbar\omega_0}{\tau_{OP}}\biggl(\frac{T_0}{\hbar\omega_0}\biggr)^2
\biggl[
\exp\biggl(\frac{\hbar\omega_0}{T_0}-\frac{\hbar\omega_0}{T^{\mp}}\biggr) - 1\nonumber\\
 + c\frac{(T^{\mp})^3-T_0^3}{T_0^3}\biggr].
\end{eqnarray}

Here, 
 $ \tau_{OP}= \tau_{0}(T_0/\hbar\omega_0)^2 \exp(\hbar\omega_0/T_0)$ 
is  the "warm" carrier energy relaxation time  at $ T^{\mp}$ close to the lattice temperature $ T_0$, $\omega_0$ and  
 $\tau_{0}$ are the optical phonon energy in GMRs and  the characteristic time of the spontaneous optical phonon emission, respectively, $c = (\tau_{OP}/\tau_{SC})(\hbar\omega_0/T_0)$ 
 with $\tau_{SC}$ being the carrier energy relaxation time associated with the SC mechanism. 
When  $ T \simeq T_0$, Eqs.~(\ref{eq4}) 
transform to more standard form:
  $ R^{\mp} \propto (T^{\mp}-T_0)$.

 Equations~(\ref{eq1}) - (\ref{eq4}) lead to
  the following equations parametrically relating $j_{GMR}^{\mp}$ and $V_G$ via $T^{\mp}$ as  parameters:

\begin{eqnarray}\label{eq5}
j_{GMR}^{\mp} \simeq j_N\frac{V_G\mp V_A}{V_G}
\biggl[\exp\bigg(\frac{\hbar\omega_0}{T_0}- \frac{\hbar\omega_0}{ T^{\mp}}\biggr)-1\nonumber\\
 +c\frac{(T^{\mp})^3-T_0^3}{T_0^3}\biggr]
\end{eqnarray}
and
\begin{eqnarray}\label{eq6}
j_{GMR}^{\mp} = 
j_T\biggl\{\frac{T^{\mp}}{T_0}\exp\biggl[
\frac{e\sqrt{{\overline V_G}(V_G\mp V_A)}+eV_G/2
-\Delta }{T^{\mp}}\biggr]\nonumber\\ - 
\frac{T^{\pm}}{T_0}\exp\biggl[
\frac{e\sqrt{{\overline V_G}(V_G \mp V_A)}-eV_G/2
-\Delta }{T^{\pm}}\biggr]\biggr\}.
\end{eqnarray}

Combining Eqs.~(\ref{eq5}) and (\ref{eq6}), we arrive at the following equation, which relates the carrier temperatures $T^{-}$ and $T^{+}$ in the n-GNRs and p-GNRs and the bias voltages $V_G$:

\begin{eqnarray}\label{eq7}
\beta_N 
\frac{(V_G\mp V_A)}{V_G}\biggl[\exp\bigg(\frac{\hbar\omega_0}{T_0}- \frac{\hbar\omega_0}{ T^{\mp}}\biggr)-1+ c\frac{(T^{\mp})^3-T_0^3}{T_0^3}\biggr]\nonumber\\
 =
\frac{T^{\pm}}{T_0}
\exp\biggl[\frac{e\sqrt{{\overline V_G}(V_G\pm V_A)}+eV_G/2-\Delta }{T^{\pm}}\biggr]
\nonumber\\ 
- 
\frac{T^{\mp}}{T_0}
\exp\biggl[\frac{e\sqrt{{\overline V_G}(V_G\pm V_A)}-eV_G/2-\Delta }{T^{\mp}}\biggr],
\end{eqnarray}
where
$\beta_N= [\pi\,HC_G\,\hbar/2(2N-1)e^2\tau_0^{\varepsilon}] (T_0/\hbar\omega_0)$.
If $H = 1.0~\mu$m, $C_G =1$, and $\tau_{OP} = 20$~ps, 
$\beta_N \simeq 4.268\times 10^{-3}/(2N-1)$.

If $V_A =0$, $T^{\mp} = T$ and Eqs.~(\ref{eq7})  yield

\begin{eqnarray}\label{eq8}
\frac{\beta_N}{2}\biggl[\exp\bigg(\frac{\hbar\omega_0}{T_0}- \frac{\hbar\omega_0}{ T}\biggr)-1+ c\frac{T^3-T_0^3}{T_0^3}\biggr]\nonumber\\
=\frac{T}{T_0}
\exp\biggl(\frac{e\sqrt{{\overline V_G}V_G}-\Delta }{T}\biggr)
\sinh\biggl(\frac{eV_G}{2T}\biggr).
\end{eqnarray}

Equation~(\ref{eq8}) leads to the $T-V_G$ characteristics, which can exhibit the thermal breakdown provided sufficiently small values of $\beta_N$ [or large number of the
GNR bridges $(2N-1)$] predicted previously.~\cite{28}

The net terminal current in the PGM with $M$ GMR pairs is equal to

\begin{eqnarray}\label{eq9}
J_{GMR} = M (2N-1)(j_{GMR}^{-} + j_{GMR}^{+}).
\end{eqnarray}

\begin{table*}[t]
\centering 
\vspace{2mm}
\begin{tabular}{|c|c|c|c|c|c|c|c|c|c|c|c|c|c}
\hline
 $H$, nm& $L$, nm &
  $L_G$, nm& $w$, nm& 
 $\Delta$, meV& $\kappa_S$&
 $C_G$ & ${\overline V}_G$, mW& 
 $T_0$, meV& $\hbar\omega_0$, meV& 
 $\tau_{OP}$, ps & $\tau_{SC}$, ps& 
 $\beta_1$\\
\hline
1000 & 25 & 25 & 10 & 250 & 10 & 1 & 8 & 25&200&20&160&4,27$\times 10^{-3}$ \\
\hline
\end{tabular}
\caption{\label{table} Main PGM parameters} 
\end{table*}

\section{Results: Voltage Dependences of Carrier Effective Temperatures
and Current-Voltage Characteristics }

\begin{figure}[b]
\centering
\includegraphics[width=7.0cm]{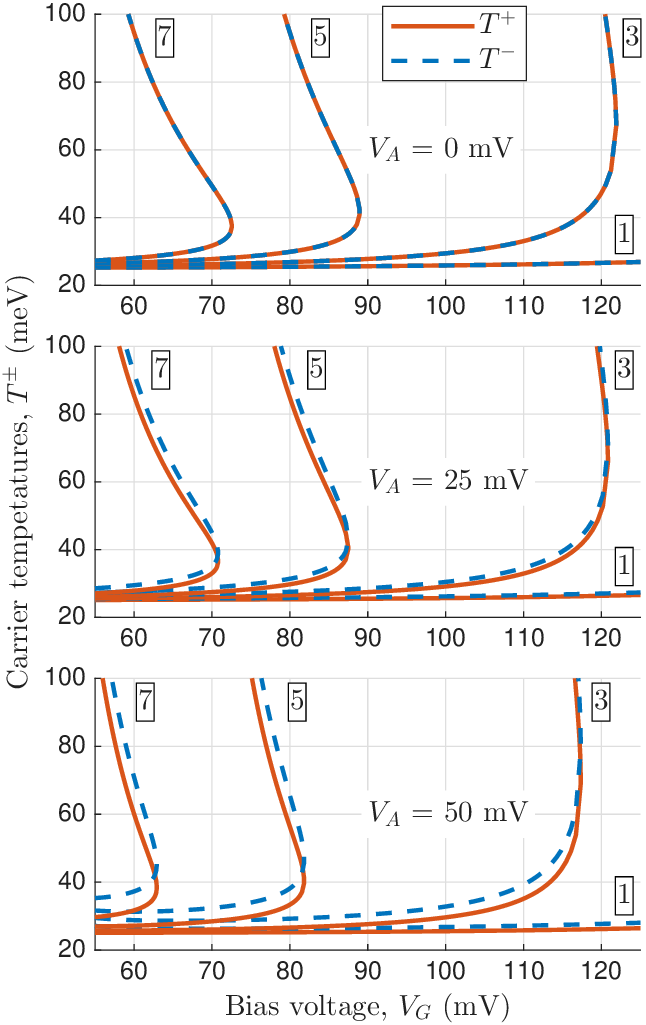}
\caption{Voltage dependences of carrier temperatures  $T^-$ and $T^+$ in the  PGMs with different  numbers, $(2N-1) =1,\, 3,\, 5,\, 7$ of the GNR bridges (indicated in rectangles) and different  doping voltages: $V_A = 0$ ($T^-=T^+ = T$) ,
 $V_A = 25$~mV, and  $V_A = 50$~mV (acceptor surface densities $\Sigma_A =0$, $\Sigma_A \simeq 3.47\times 10^{11}$~cm$^2$, and  $\Sigma_A \simeq 6.94\times 10^{11}$~cm$^2$).}
\label{Fig2}
\end{figure}

\begin{figure}[t]
\centering
\includegraphics[width=7.0cm]{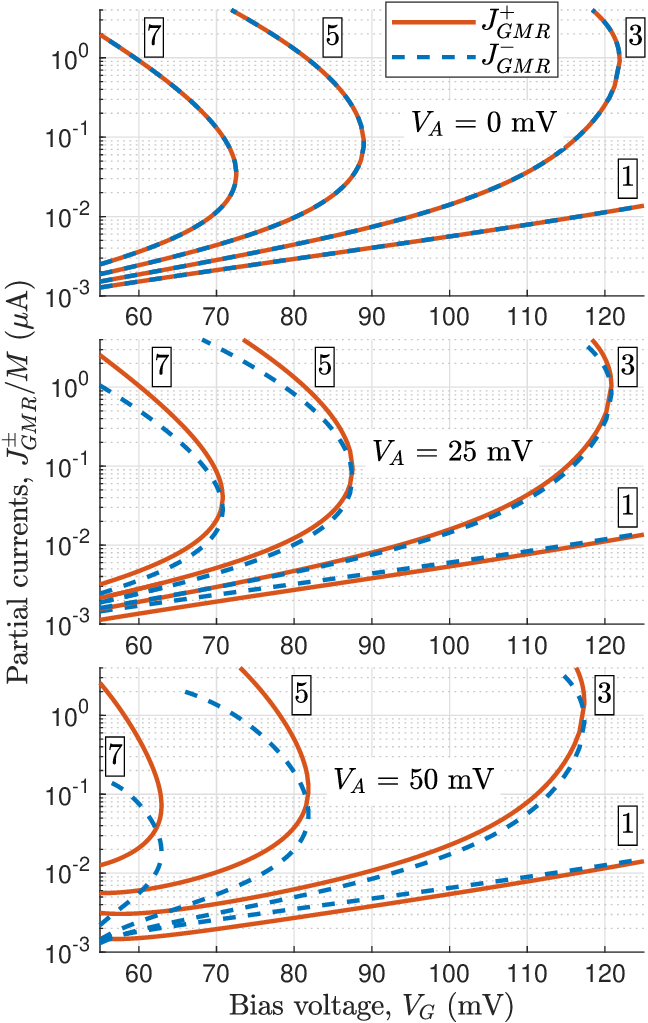}
\caption{Voltage dependences of partial currents of $J_{GMR}^-/M$ and $J_{GMR}^+/M$ normalized by the number, $M$, of the GMR pairs   in the  PGMs with different numbers, $(2N-1)$,  of the GNR bridges and different  doping voltages $V_A$.}
\label{Fig3}
\end{figure}

\begin{figure}[t]
\centering
\includegraphics[width=7.0cm]{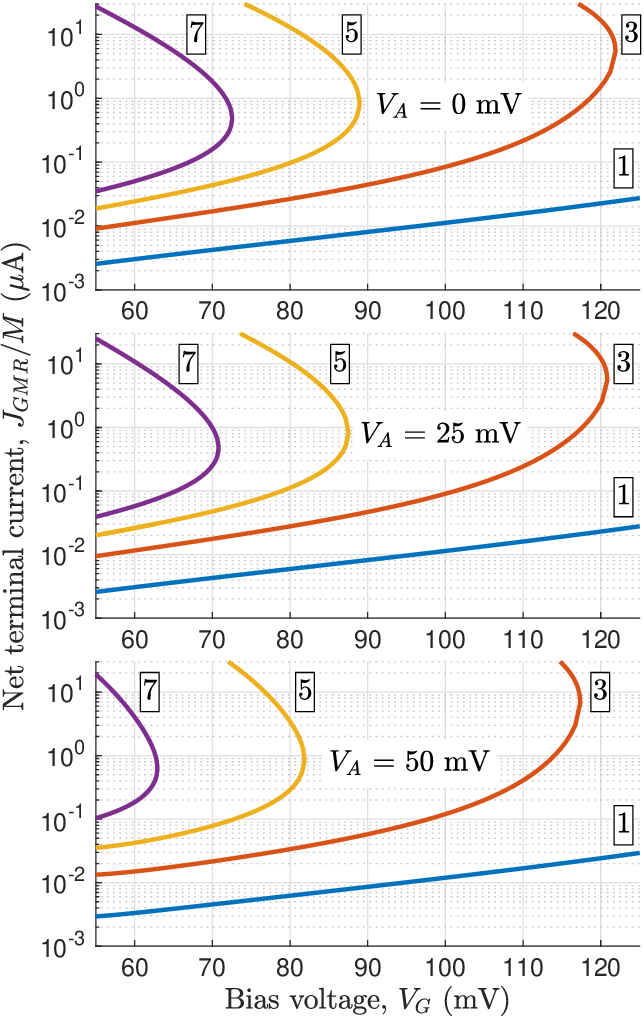}
\caption{Normalized current-voltage characteristics $J_{GMR}/M$ of  the PGMs with  different  $(2N-1)$ and $V_A$.
}
\label{Fig4}
\end{figure}

\begin{figure}[t]
\centering
\includegraphics[width=6.0cm]{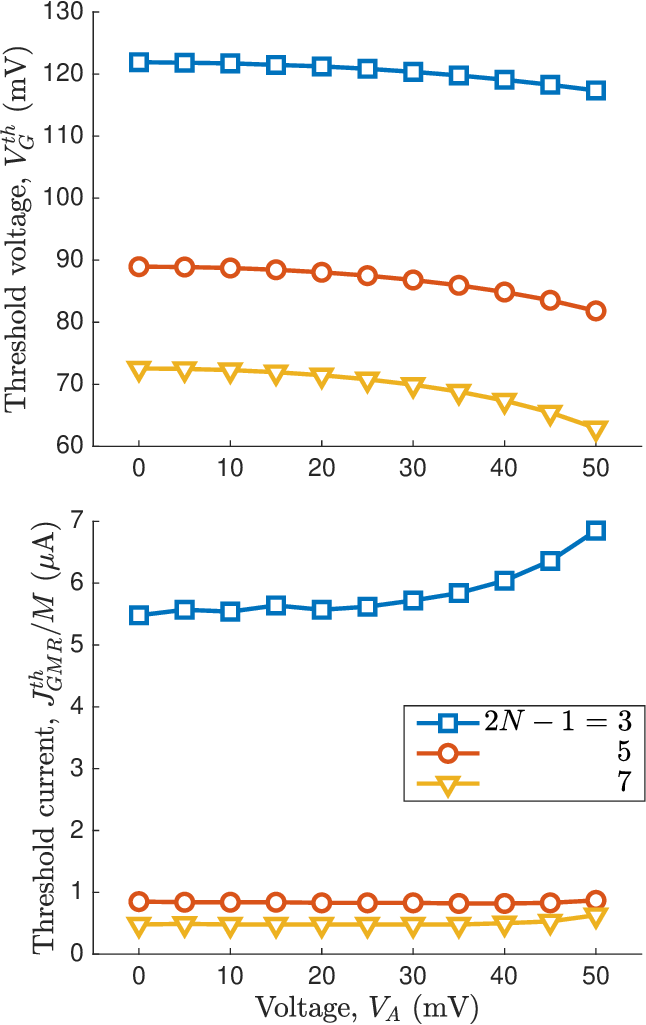}
\caption{Threshold voltage $V_G^{th}$(top panel) and threshold normalized current $J_{GMR}^{th}$  (bottom panel) as functions of  the doping voltage $V_A$ for the PGMs with different $(2N-1)$.
}
\label{Fig5}
\end{figure}

For  the illustration, we solved  the system of coupled  Eqs.~(\ref{eq7}) numerically 
using  MATLAB (R2024b). Since the pair of implicit Eqs.~(\ref{eq7}) with multiple solutions for $T$ can not be
solved directly by the exchange of variables, we utilized MATLAB built-in function {\it vpasolve}, calculating  parts of the curves
of the system below and above (on the temperature axis) the threshold points separately, guessing their initial value ranges.
The main PGM parameters used for the calculations are listed in Table~I and include a different number,
$(2N-1)$, of GNR bridges connecting the GMRs and different donor voltages $V_A$ in the range $V_G > V_A$.  The calculation results are presented in Figs.~2--4.

Figure~2(a)  shows that in the undoped PGMs ($V_A =0$),
the carrier temperatures in the neighboring GMRs are equal ($T^- = T^+ = T$
being transformed from  monotonous (rising, sublinear)  and ambiguous functions of the bias voltage $V_G$  depending on the number of the GNR bridges $(2N-1)$. 
This transformation is due to the reinforcement of the injection between the GMRs with an increasing number of injecting channels, i.e., the number of the GNR bridges. 
The latter results in a stronger carrier overheating. We refer to such a characteristic transformation as 
hot-carrier thermal breakdown. It is attributed to positive feedback between the carrier heating and 
carrier injection between neighboring GMRs.
 The ambiguous dependences
are characterized by the threshold voltage $V_G^{th}$, at which the derivative $dT/dV_G$  escapes to infinity.
Apart from two branches of the $T - V_G$ shown in Fig.~2,   the ambiguous characteristics corresponding to a sufficiently large $(2N-1)$
exhibit  the appearance of the third, high-temperature branch with $dT/dV_G > 0$.  These $T - V_G$ relations for these branches are determined by
the high-temperature mechanisms of carrier relaxation (focusing on the doping effects, we do not consider the high-temperature branch here, see Ref.~29). Hence, at large $(2N-1)$, the  $T - V_G$ relations have an S-shape. 

As seen in Fig.~2 for doped PGMs (middle and lower panels), the doping
leads to a splitting of the $T^{-} - V_G$ and $T^{+}-V_G$ voltage dependences
associated with the non-equivalence (violation of the symmetry) of the n- and p-GMRs. 
Comparing the panels in Fig.~2, one sees that
this splitting increases with increasing doping level (increasing doping voltage $V_A$) with the electron temperature $T^{-}$ somewhat exceeding the hole temperature $T^{+}$. This is because the electron density is smaller
than that of the holes (if $V_A > 0)$, while the hole current injected into the n-GMR is larger than the electron current injected into the p-GMRs (see Fig.~3 below).  
 Nevertheless, both  the $T^{-} - V_G$
and $T^{+} - V_G$ relations are characterized by the same threshold voltage $V_G^{th}$. 
 
 Figure 3  shows that the doping affects the voltage dependences of the 
 partial currents $J_{GMR}^{-}/M = (2N-1) j{GMR}^{-} $  and $J_{GMR}^{+}/M = (2N-1) j{GMR}^{+} $ (per one GMR pair),
also  resulting in their splitting,
 although the type of these characteristics is preserved (at least at the doping levels under consideration). Stronger variations of the $J_{GMR}^{-} - V_G$  and $J_{GMR}^{+} - V_G$ characteristics compared to the 
 $T^{-} -V_G$  and $T^{+} -V_G$ relations are attributed to an exponential dependence of the partial  currents  on the carrier temperatures $T^{-}$ and $T^{+}$ and the carrier Fermi energies $\mu^{-}$ and $\mu^{+}$. The $T^{-} - V_G$
and $T^{+} - V_G$ relations on the one hand and the $J_{GMR}^{-} - V_G$ and
 $J_{GMR}^{+} - V_G$ characteristics   on the other  correspond to the same values of the threshold voltage $V_{G}^{th}$.   
  The latter is because when $dT^{\pm}/dV_{G} \Rightarrow \infty$
at $V_G$ tending to $V_G^{th}$,
$dJ_{GMR}^{\pm}/dV_G \Rightarrow \infty$ , due to proportionality of $dJ_{GMR}^{\pm}/dV_{G}$ to $d T^{\pm}/dV_{G}$.
   
Although the splitting of the $T^{-} - V_G$ and  $T^{+} - V_G$ characteristics with increasing
doping voltage appears to be moderate,                      
the doping affects the 
$J_{GMR}^{-} - V_G$ and   $J_{GMR}^{+} - V_G$ characteristics 
much more pronouncedly.

 As seen in Fig.~4, the net current-voltage characteristics $J_{GMR} - V_G$
(normalized by the number, $M$, of GMR pairs) qualitatively repeat the behavior of the partial currents, being also sensitive to the doping level and, particularly, to the number of the GNR bridges $(2N-1)$. 

Figure~5 shows the noticeable (but not critical) variations in the threshold voltage and normalized threshold current, $V_G^{th}$ and $J_{GMR}^{th}/M$, with increasing  doping voltage $V_A$. 
At the same time, as can be seen from Fig.~5 (and Fig.~4), 
the former
are very sensitive to the number of the GNR bridges $(2N-1)$.

\section{Conclusions}
We have shown that: (1) depending on the PGM structural parameters,   the carrier temperature-voltage in the PGMs and their current-voltage characteristics can be fairly steep, being monotonous or ambiguous and tending to a S-shape - the effect of  hot-carrier breakdown;
(2) these dependences are  modified  with increasing doping level;
(3) the effect of hot-carrier breakdown is preserved up to rather high doping levels in the PGMs, being fairly sensitive to the number of the GNR bridges connecting the neighboring GMRs. 

The obtained results can be used for the realization of novel PGM-based
devices, such as fast voltage-controlled switches of the current, thermal incandescent
light sources, and terahertz bolometers  and their optimization.

\section*{ACKNOWLEDGMENTS}
\vspace{-5mm}
The work at Research Institute of Electrical Communication
(RIEC), Frontier Institute for Interdisciplinary Studies
(FRIIS), and University of Aizu (UoA) was supported
by the Japan Society for the  Promotion of Science (KAKENHI
Grants Nos. 25K01281 and 25K22093), 
The Telecommunications
Advancement Foundation, SCAT,
 JST ALCA-Next, and NEDO, Japan.
 The work at ENSEMBLE3 Ltd. was supported under
the International Research Agenda program of the Foundation
for Polish Science (FENG.02.01-IP.05-0044/24),
 the European
Union through the Horizon 2020 Teaming for Excellence
program (GA No. 857543), and 
the Poland Minister of Science
and Higher Education Center of Excellence project under
the Horizon 2020 program (No. MEiN/2023/DIR/3797).


\section*{Data availability}
\vspace{-5mm}
All data that support the findings of this study are available
within the article.

\end{document}